\documentclass{IEEEtran}
\usepackage[english]{babel}
\usepackage[T1]{fontenc}
\usepackage{cite}
\usepackage{amsmath,amssymb,amsfonts}
\usepackage{graphicx}
\usepackage[dvipsnames]{xcolor}
\usepackage{textcomp,nicefrac}
\usepackage{verbatim}
\usepackage{bm}
\usepackage{siunitx} 
\DeclareSIUnit\sq{\ensuremath{\Box}}                           

\usepackage{booktabs} 
\usepackage{multirow} 
\usepackage{subcaption}

\newcommand\txtc{black}

\def\BibTeX{{\rm B\kern-.05em{\sc i\kern-.025em b}\kern-.08em
T\kern-.1667em\lower.7ex\hbox{E}\kern-.125emX}}
\begin{document}
\title{ MiniCACTUS: A 65 ps Time Resolution Depleted Monolithic \textcolor{\txtc}{CMOS} Sensor}
\author{Yavuz Degerli, Fabrice Guilloux, Tomasz Hemperek, Jean-Pierre Meyer, and Philippe Schwemling}
\thanks{Manuscript received July 3, 2023,; revised September X, 2023. 
This project has received funding from the European Union’s Innovation programme under grant no 101004761 (AIDAInnova). \textit{(Corresponding author: Yavuz Degerli)}.}
\thanks{Yavuz Degerli is with IRFU/DEDIP, CEA-Saclay, France (e-mail: yavuz.degerli@cea.fr).}
\thanks{Fabrice Guilloux is with IRFU/DEDIP, CEA-Saclay, France (e-mail: fabrice.guilloux@cea.fr}
\thanks{Tomasz Hemperek was with the Physics Department, University Bonn. He is now with Dectris, Switzerland}
\thanks{Jean-Pierre Meyer is with IRFU/DPhP, CEA-Saclay, France (e-mail: jpmeyer@cea.fr).}
\thanks{Philippe Schwemling is with IRFU/DPhP, CEA-Saclay and Universit\'e Paris-Cit\'e, France (e-mail: philippe.schwemling@cea.fr).}

\maketitle

\begin{abstract}

MiniCACTUS is a monolithic sensor prototype optimised for timing measurement of charged particles. It has been designed in a standard 150\,nm CMOS process without dedicated amplification layer.
It is intended as a demonstrator chip for future large scale timing detectors, like upgrades of timing detectors at LHC, or future high energy physics detector projects. The sensor features an active array of \qtyproduct{2 x 4}{}~diodes, analog and digital Front-Ends (FEs), a slow control interface, and bias circuitry programmable through internal DACs. The sensing element is a deep n-well/p-substrate diode. Thanks to the optimized guard-rings surrounding the whole chip, it is possible to apply safely more than -\,450\,V on the high-resistivity substrate allowing fast charge collection. The baseline pixel dimensions are \textcolor{\txtc}{1.0\,mm$\times$ 1.0\,mm and 0.5\,mm$\times$ 1.0\,mm}. The analog FEs and the discriminators for each pixel are implemented outside the pixel, at the column level. The power consumption \textcolor{\txtc}{is approximately 300\,mW/cm{$\mathbf {^2}$}}, which is compatible with \textcolor{\txtc}{cooling infrastructure} available at LHC experiments, and making integration of this concept viable in future high energy physics experiments. After fabrication, the sensors have been thinned to 100\,µm, 200\,µm and 300\,µm total thickness and then post-processed for backside biasing. The time resolution of several sensors with different thicknesses has been measured in 3 test-beam campaigns using high energy muons (Minimum Ionizing Particles) at CERN SPS in 2021 and 2022. A resolution of 65.3\,ps has been measured with on-chip FE and discriminator. This paper will focus on the results of these test-beam campaigns.

\end{abstract}

\begin{IEEEkeywords}
CMOS sensor, monolithic active pixel sensors (MAPS), Timing sensor 
\end{IEEEkeywords}

\section{Introduction}
The measurement of the time of arrival of charged particles in high energy physics experiments is attracting significant interest, in existing experiments as well as in future experimental projects. Both ATLAS and CMS collaborations at LHC are planning the installation of dedicated timing oriented subdetectors, the HGTD~\cite{HGTD} in the case of ATLAS, the MTD~\cite{MTD} and HGCAL~\cite{HGCAL} in the case of CMS. The sensors on which these new subdetectors are based are Low Gain Avalanche Diodes (LGADs)~\cite{Apresyan}. Detector projects whose time scales are longer than LHC Phase 2 upgrades, like EIC, or FCC-ee/hh, are all seriously considering the implementation of timing detectors with a resolution better than \SI{100}{ps}. The motivation for such measurements is the need to complement standard tracking momentum and position measurements with additional information to help disentangle multiple interactions in heavily crowded environments, induced by a high pile up level. An additional motivation is low energy particle identification through time of flight measurement.

Technical requirements for timing detectors are often difficult to meet and sometimes contradictory. For hadronic environments, radiation hardness is the main difficulty. Another \textcolor{\txtc}{constraint} is the need for relatively low power consumption (no more than a few hundreds of \SI{}{\milli\watt\per\centi\meter\squared}). This low power is mandatory to ensure the thermal stability of the detector, and to minimize the amount of dead material brought in by the cooling system, but it comes in conflict with time resolution performance, since in general the higher the power, the faster the analog processing can be, hence enhancing time resolution. \textcolor{\txtc}{Typical detector concepts where these constraints have to be taken into account are detailed in~\cite{ibl-tdr} and~\cite{itk-tdr}.}

Among the many approaches that are explored today, CMOS Depleted Monolithic Pixels Sensors present several appealing features compared with hybrid solutions. They can be designed using intrinsically radiation hard standard High Voltage (HV) technologies on High Resistivity (HR) wafers, like LFoundry LF15A, where the charges \textcolor{\txtc}{are collected} by drift and not only by diffusion. Since the Front-End (FE) and readout electronics reside on the same substrate as the sensor diode, there is no need for a costly and delicate bump bonding operation~\cite{Peric}. The foundry processes used are \textcolor{\txtc}{high production volume}, potentially allowing for very reproducible performance from sensor to sensor, and low cost when produced in high volumes of several \SI{}{\meter\squared}, as will be needed by future collider experiments. In addition, LF15A process has been successfully evaluated for tracking applications in heavily ionizing environments like HL-LHC, with a suitable signal to noise ratio up to doses of \SI{150}{Mrad} \cite{Chen,Barbero}.

The MiniCACTUS presented in this paper is a Monolithic Depleted CMOS sensor prototype implemented in LF15A \SI{150}{nm} technology, to investigate the possibility of time measurement with a \SI{100}{ps} or better resolution on Minimum Ionizing Particles (MIP), relying only on the collection of the charge deposited in the silicon, without intrinsic amplification.
\label{sec:introduction}

\section{Sensor description}
\begin{figure}[htbp]
\centerline{\includegraphics[width=1.8in]{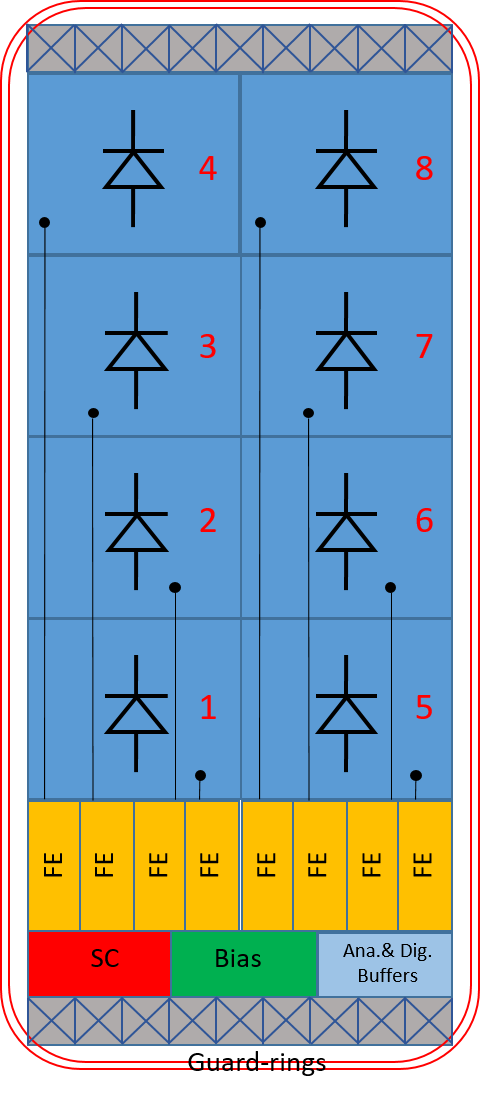}}
\caption{Architecture of the MiniCACTUS sensor prototype chip (not to scale).}
\label{minicactus}
\end{figure}

\begin{figure}[htbp]
\begin{subfigure}{\linewidth}
    \centerline{\includegraphics[width=3.0in]{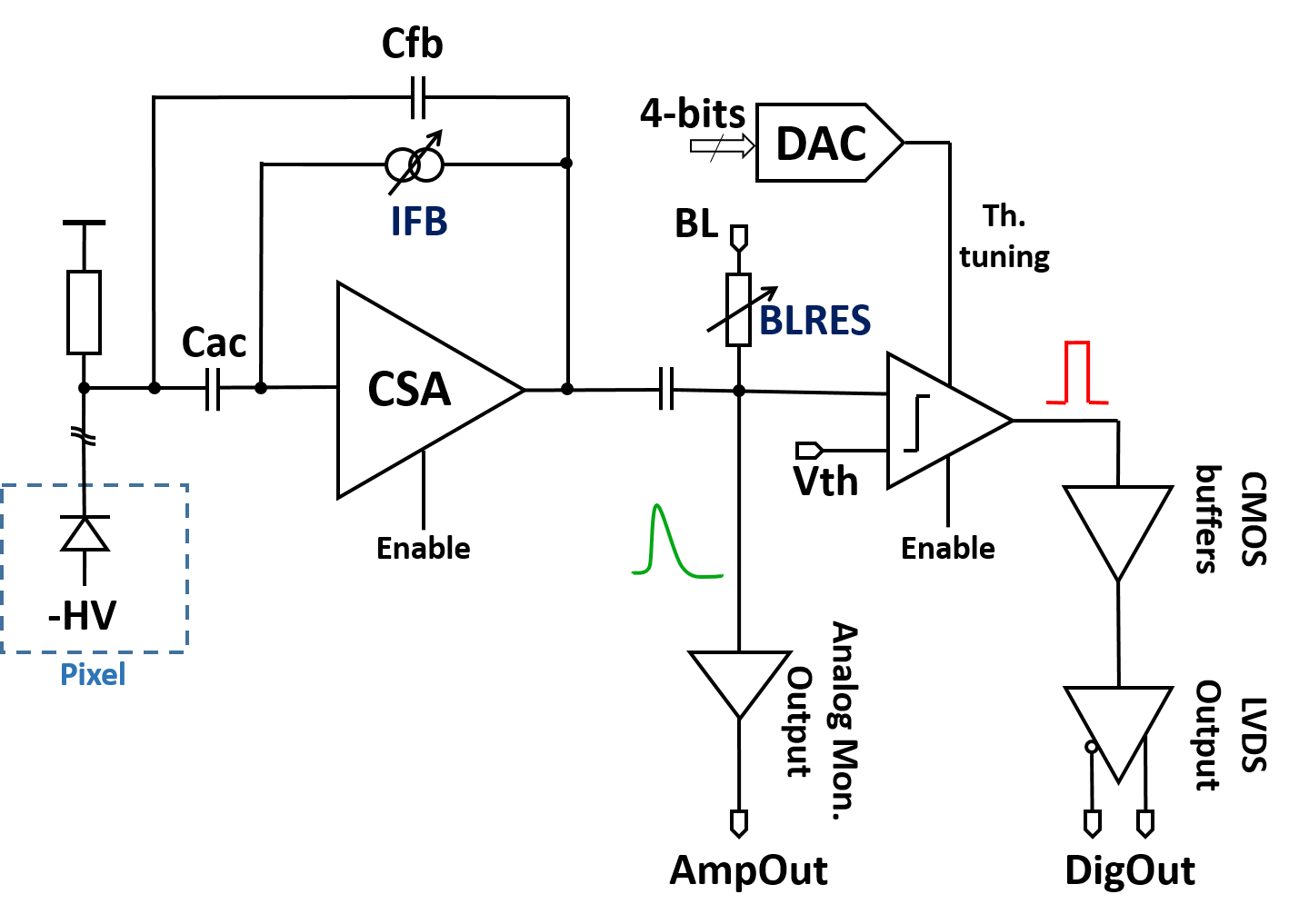}}
    \caption{}
\end{subfigure}
\begin{subfigure}{0.5\textwidth}
    \centerline{\includegraphics[width=3.0in]{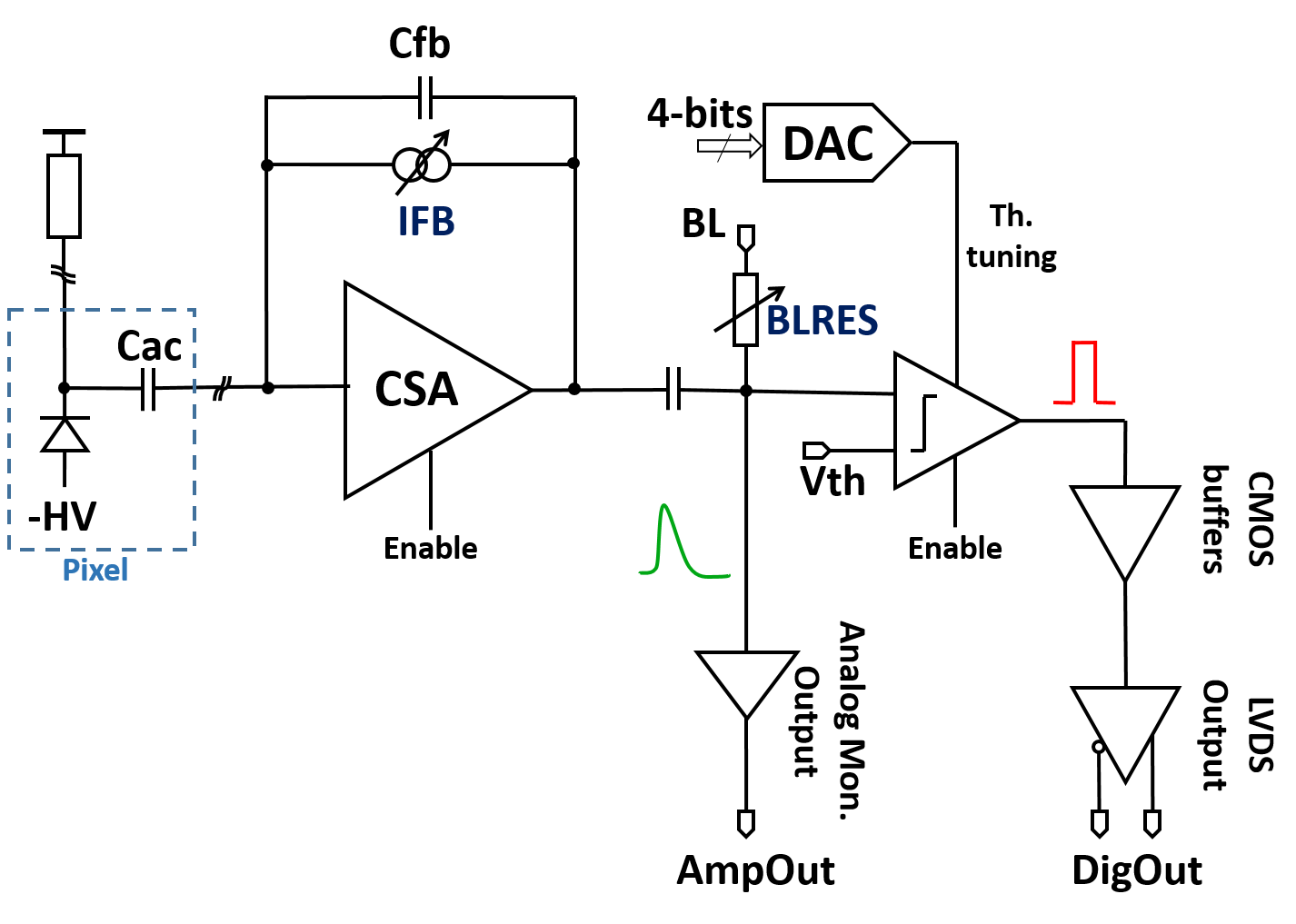}}
    \caption{}
\end{subfigure}
\caption{MiniCACTUS Front-End architecture. \textcolor{\txtc}{a) Pixels 1-7:  the AC coupling capacitor Cac is integrated into the feedback loop, b) Pixel 8: the AC coupling capacitor is integrated on top of the diode.}}
\label{FE}
\end{figure}

The MiniCACTUS sensor chip visible in Fig.~\ref{minicactus} has been designed and fabricated in the LF15A \SI{150}{nm} process. The chip size is \qtyproduct{3.5 x 2.5}{\milli\meter}. It comprises mainly an array of 8 passive diodes with associated column-level FE circuitry for each one. \textcolor{\txtc}{In previous CACTUS sensors, the FE has been implemented inside the pixel \cite{cactus}. Nevertheless, the metal rails needed to distribute the FE power over large pixels in Y-direction add significant parasitic to the detector capacitance. In order to address this issue, in MiniCACTUS the FE has been implemented outside the pixel, at the column-level. Putting the FE inside the collecting diode is technically feasible but needs a revised power lines network strategy to not add additional parasitic capacitances and was not studied in this work.}
The FE of MiniCACTUS is based on an AC-coupled Charge Sensitive Amplifier (CSA), a leading edge discriminator, and a 4-bit DAC for threshold tuning (see Fig.~\ref{FE}). \textcolor{\txtc}{It does not include a leakage compensation circuitry. For the baseline pixels (1\,--\,7), the AC coupling capacitor Cac is implemented inside the feedback loop. With this configuration, this capacitor does not need to be larger than the sensor capacitance \cite{Peric}}. The charge collection diode structure, which is basically a deep N-Well in a p-type high resistivity substrate, has been described in \cite{minicactus}. The FE has been designed for a peaking time of about \SI{1}{ns} and for a detector capacitance of \SI{1.5}{pF}, matching the results obtained from initial TCAD simulations \cite{cactus}.

The chip has analog and digital multiplexing circuitry for the selection of single pixel readout, output drivers, and a common bias circuitry block with global 6-bit DACs. A 204-bit shift register is used as SPI-like slow-control (SC), allowing single FE bias parameters programming and enabling/disabling the amplifiers and discriminators for each pixel individually. The outputs of the discriminators are sent to output pads via standard CMOS buffers. LVDS translators/drivers, integrated inside the pads, are used to send the digital signals out of the chip. It is possible to monitor the output of the CSA for a chosen pixel (only 1 pixel per column at a time). Due to the limited bandwidth of the buffer (less than 1 GHz), this output is not an exact copy of the analog signal.

\begin{figure}[htbp]
\centerline{\includegraphics[width=3.2in]{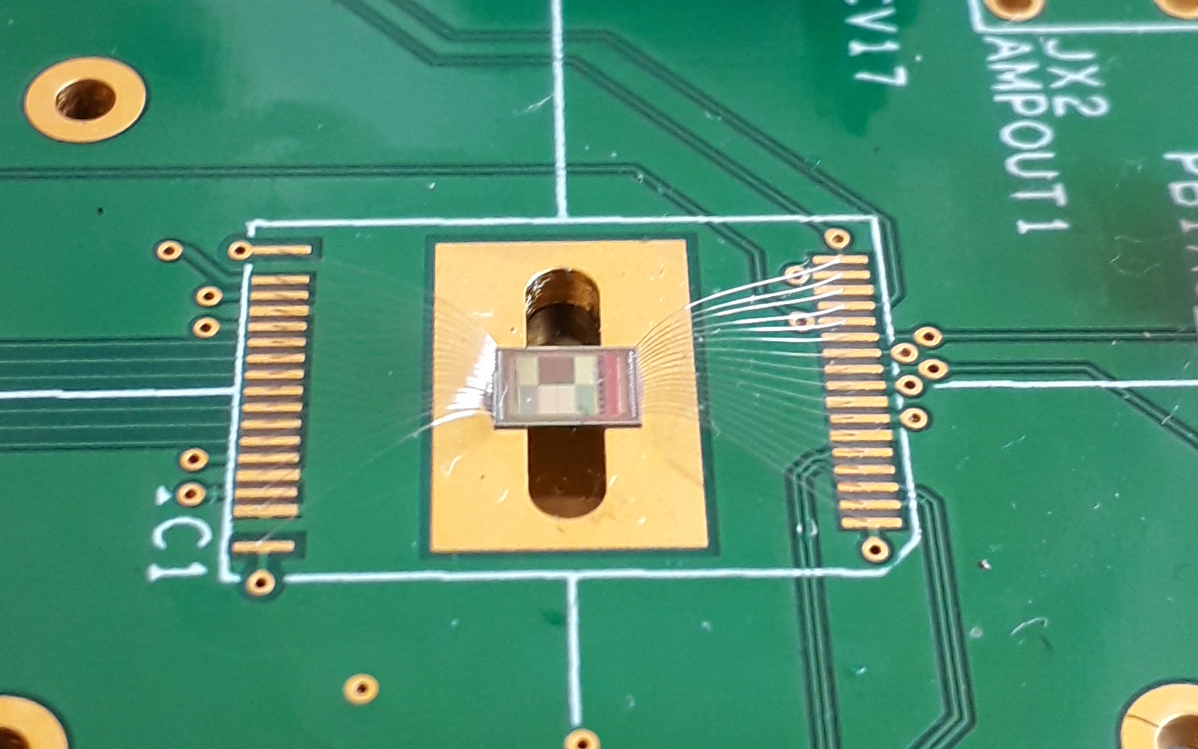}}
\caption{Picture of a MiniCACTUS sensor bonded on a PCB. The chip size is \qtyproduct{3.5 x 2.5}{\milli\meter}.}
\label{pcb-minicactus}
\end{figure}

All pixels in MiniCACTUS are passive. One pixel among eight, pixel 8 (Fig. ~\ref{minicactus}), has its large Metal-on-Metal AC coupling capacitance Cac integrated on the top of the diode (Fig. ~\ref{FE}b). The baseline pixel pitches are \qtyproduct{1 x 1}{\milli\meter} (pixels 3 and 7) and \qtyproduct{0.5 x 1}{\milli\meter} (pixels 2, 4, 6, and 8). The nominal powers dissipation of the FE of these pixels are respectively \SI{150}{\milli W \per \centi m^2} and \SI{300}{\milli W \per \centi m^2}. There are also two small size pixels, pixel 1 (\qtyproduct{50 x 50}{\micro\meter}) and pixel 5 (\qtyproduct{50 x 150}{\micro\meter}), implemented as test structures without power optimized FE.

After fabrication, the standard HR wafers have been thinned to \SI{100}{\micro m}, \SI{200}{\micro m} and \SI{300}{\micro m} (total thicknesses), and post processed for backside polarization (p+ implantation and metallization), essential for homogeneous charge collection in timing applications. \textcolor{\txtc}{The thickness of the top metallization and passivation of the wafers is estimated to be \SI{\sim 15}{\micro m}. So, the effective thicknesses of the active part of these sensors are \SI{85}{\micro m}, \SI{185}{\micro m} and \SI{285}{\micro m}, respectively. In the rest of the paper, these numbers will be used as active sensor thicknesses.} The resistivity of the wafers used is typically \SI{2}{\kilo \Omega \cdot \centi m}.

The external guard-rings used to apply the high voltage to the substrate are the same as the ones used for CACTUS sensors~\cite{cactus}. Very similar guard-rings with minor modifications have also been used for LF-MONOPIX2 tracking sensors developed in the same process \cite{lfmonopix2}. With this guard ring geometry, it is possible to apply up to \SI{-500}{V} bias voltage to the diode of a \SI{185}{\micro m} sensor. At this bias voltage, the sensor is expected to be fully depleted over its complete thickness. \textcolor{\txtc}{In fact, according to \cite{Mandic}, for a \SI{185}{\micro m} active thickness sensor, the substrate is already fully depleted for a bias voltage of \SI{-80}{V}.}

Fig.~\ref{pcb-minicactus} shows the picture of a thinned sensor bonded on a PCB. 
Before the testbeam, the MiniCACTUS sensors have been tested extensively in laboratory conditions using $^{55}$Fe, $^{241}$Am, and $^{90}$Sr sources. The detailed results have been reported in \cite{minicactus}. Among large pixels, the best SNR is observed on
the pixel 8 -- that has an AC-coupling capacitor Cac integrated inside the pixel -- with \SI{\sim 180}{e^-} input referred noise for a \SI{185}{\micro m}-thick sensor. The same pixel exhibits a SNR degraded for \SI{85}{\micro m}-thick sensors due to the reduced signal level and increased detector capacitance. Small pixels (1 and 5) have a better SNR than large pixels thanks to the diode capacitance being smaller due to the reduced surface but these good results will not translate in a better time resolution with MIPs (detailed in section~\ref{Analysis_of_results}, see Table~\ref{Chip5-SNR} and Table~\ref{Chip3-SNR}).

\begin{figure}[htbp]

\begin{subfigure}{0.5\textwidth}
    \centerline{\includegraphics[width=3.8in]{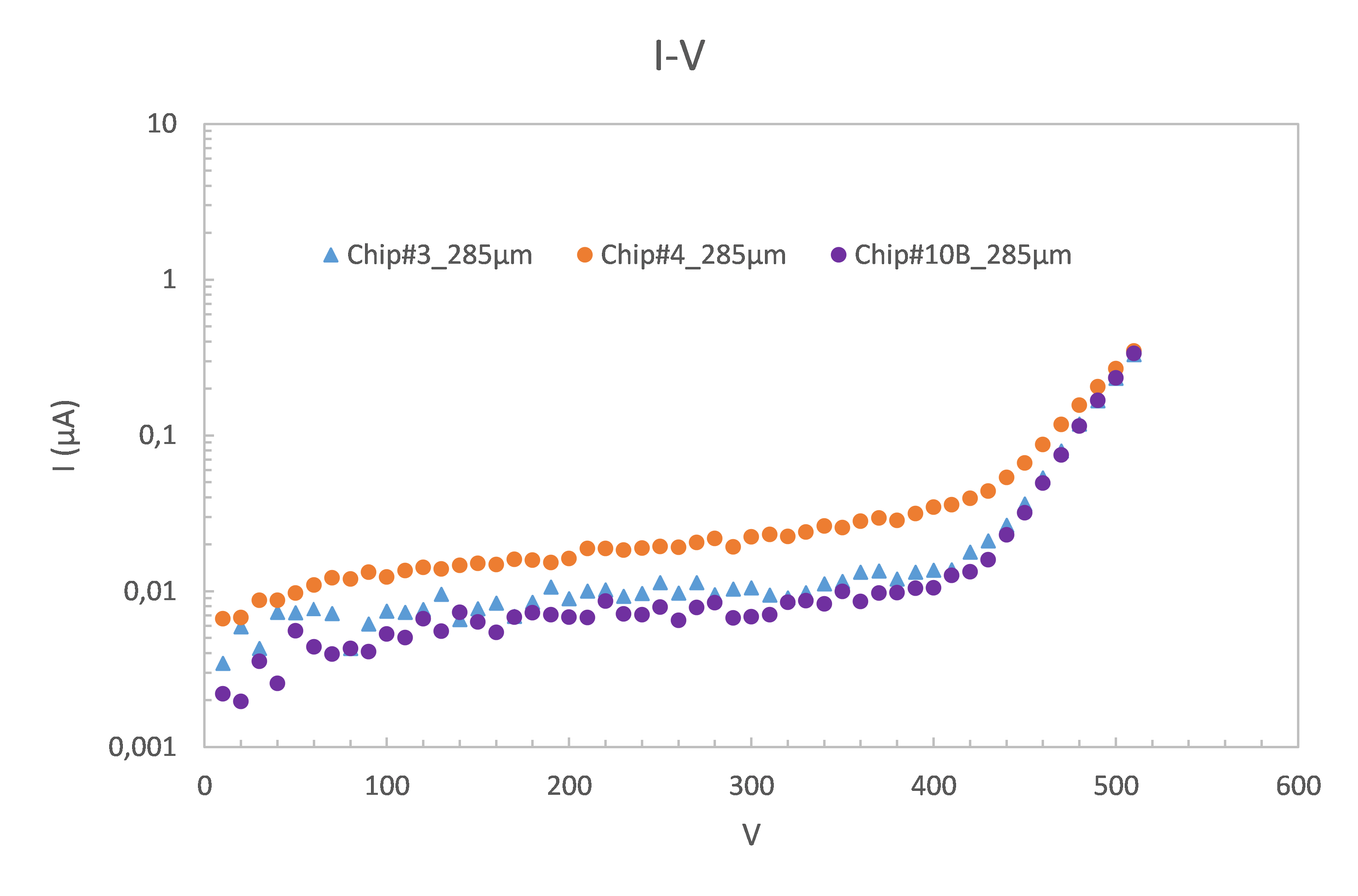}}
    \caption{}
\end{subfigure}
\begin{subfigure}{0.5\textwidth}
    \centerline{\includegraphics[width=3.8in]{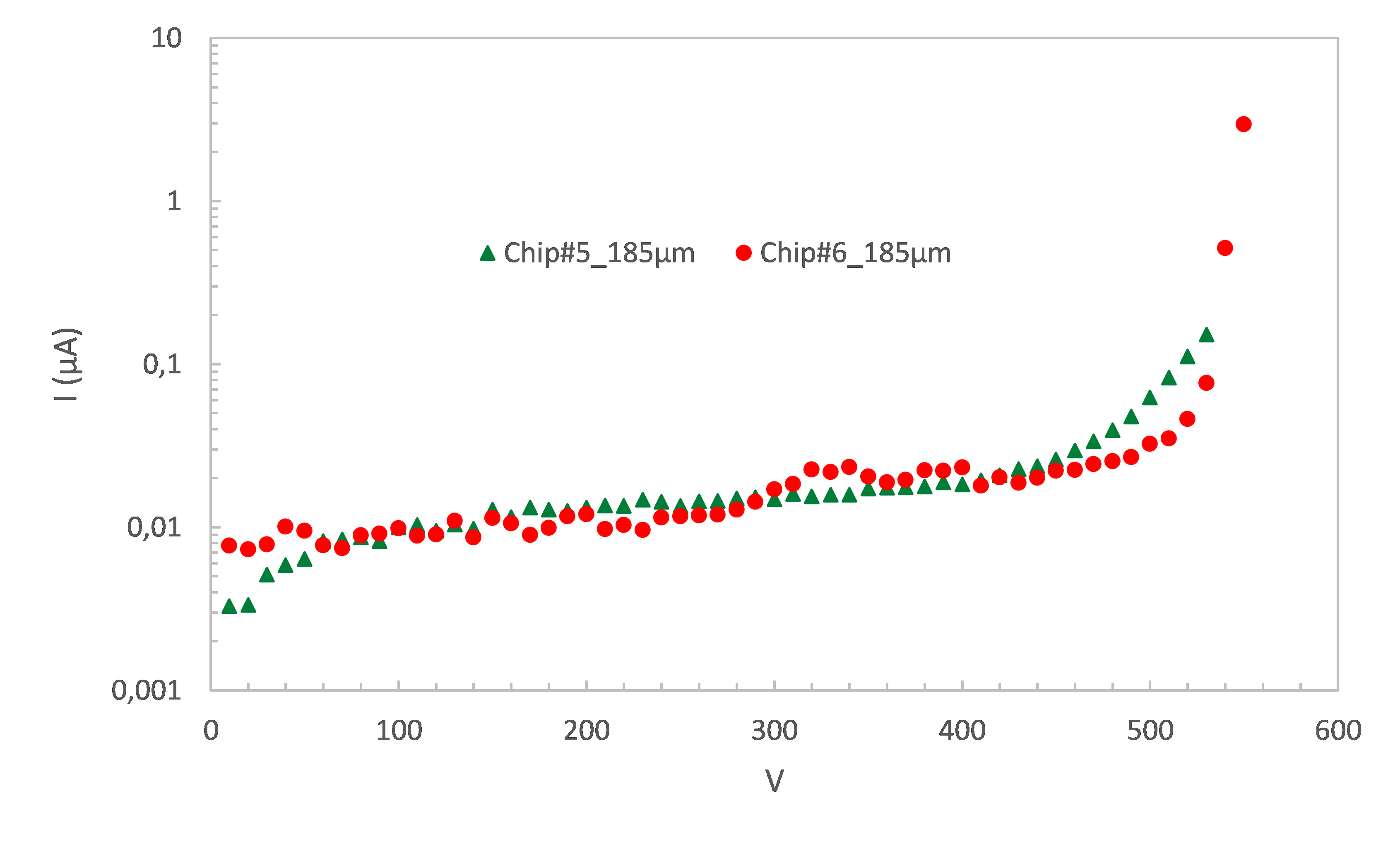}}
    \caption{}
\end{subfigure}
\begin{subfigure}{0.5\textwidth}
\centerline{\includegraphics[width=3.8in]{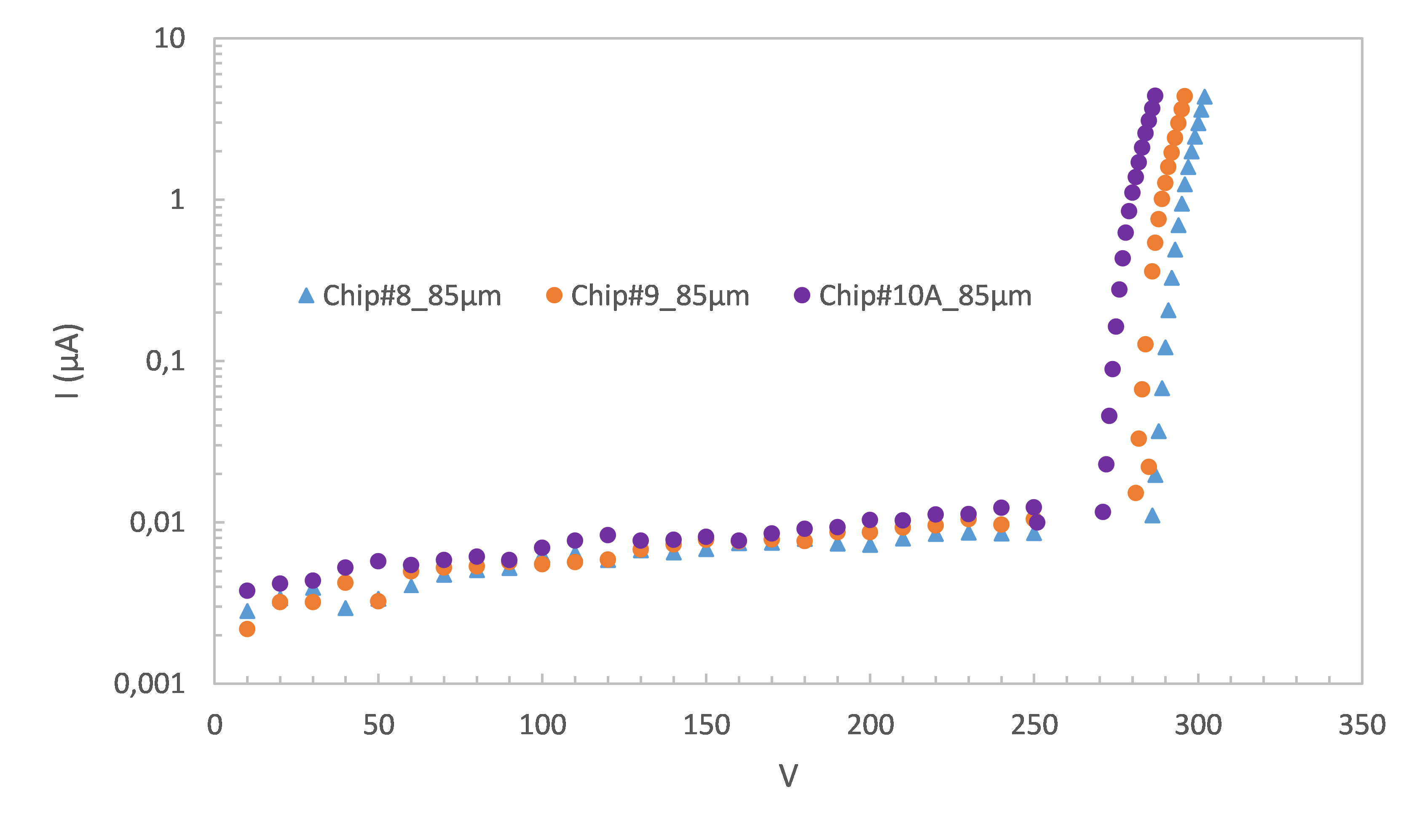}}
    \caption{}
\end{subfigure}

\caption{I-V characteristics measured on different MiniCACTUS sensors thinned to (a) \SI{285}{\micro m}, (b) \SI{185}{\micro m} and (c) \SI{85}{\micro m} with back-sides post-processed.}
\label{minicactus-IV}
\end{figure}

\textcolor{\txtc}{The I-V curves measured on 8 sensors with 3 different thicknesses are shown in Fig.~\ref{minicactus-IV}.} \textcolor{\txtc}{The measurements were done at room temperature, in a non-temperature controlled environment.} The measured breakdown voltages give the possibility to bias the substrates safely at least at \SI{-450}{V} for \SI{185}{\micro m}-thick and \SI{285}{\micro m}-thick devices.
\textcolor{\txtc}{The smaller breakdown voltage measured on \SI{85}{\micro m}-thick devices is attributed to the post-processing quality, since similar breakdown voltages have been observed on previous CACTUS devices with various thicknesses (not published). Note that, these measurements have been performed mainly in order to check the breakdown voltages. The measured absolute leakage current values are probably overestimated due to the environment conditions, like external light conditions.}

\begin{figure}[htbp]
\centerline{\includegraphics[width=2.5in]{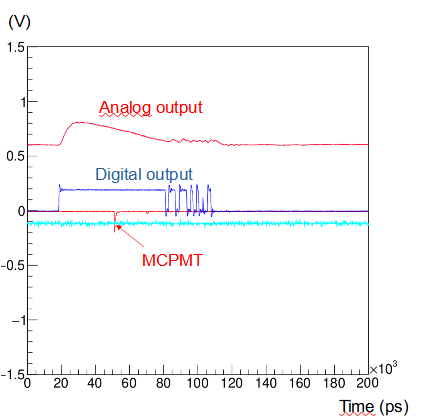}}
\caption{Typical scope traces obtained from testbeam data showing MCPPMT signal, analog (AmpOut) and digital outputs (DigOut) of the sensor.}
\label{MiniCactus-coupling}
\end{figure}

During in-lab tests, an important coupling from on-chip digital buffers into the sensors has been observed. This effect is visible on the analog monitoring and digital outputs of large pixels (\SI{1.0}{\milli\metre\squared} and \SI{0.5}{\milli\metre\squared}) in the form of ringing (see Fig. \ref{MiniCactus-coupling}). The ringing comes from digital buffers sized to drive \SI{\sim 2.35}{\milli\metre} on-chip metal lines, the distance between the outputs of the discriminators and the output pads. The single-ended CMOS buffers inject parasitic charges during transitions of discriminators. The exact position of the faulty buffers was highlighted by reducing specifically their power supply, leading to a strong attenuation of the ringing. Fortunately, the ringing affects mainly the falling edges of the analog and digital signals and does not prevent to do precise timing measurements for sensors with \SI{185}{\micro m} and \SI{285}{\micro m} thicknesses. This effect is more pronounced on \SI{85}{\micro m} sensors.

\section{Test-beam setup}
\begin{figure}[htbp]
\centerline{\includegraphics[width=2.5in]{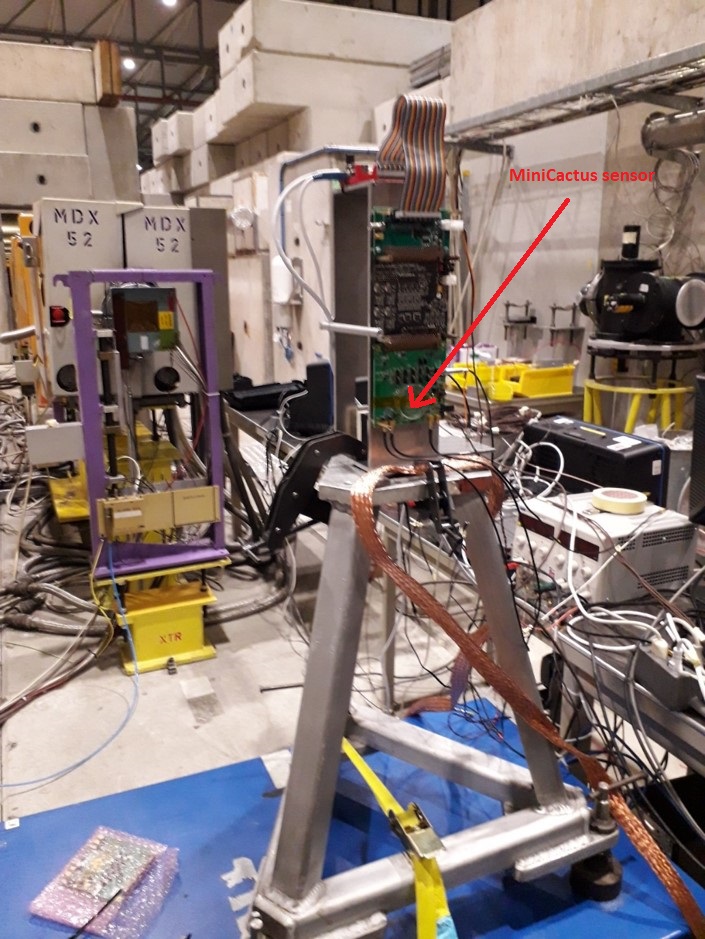}}
\caption{Testbeam setup. The sensor itself is indicated by the red arrow. The board holding the sensor is attached to the GPAC board providing slow control connection and control voltages. Towards the left of the photograph the beamline is visible in the background.}
\label{photo-tbeam}
\end{figure}

Six different MiniCACTUS sensors have been tested at CERN, in the North area, on the H4 beamline during 3 campaigns in 2021 and 2022. The tests have been done together with the RD-51 collaboration. The RD-51 testbeam setup features a tracker, which was used to align the MiniCACTUS with the beam center, and several \textcolor{\txtc}{Micro Channel Plate Photomultiplier Tubes (MCPPMTs)} with an intrinsic time resolution better than \SI{10}{ps} \cite{mcp}. The timing performance measurements of the MiniCACTUS did not use the tracker information. The analog signal from one of the MCPPMT was used as a time reference for MiniCACTUS. Fig.~\ref{photo-tbeam} shows a picture of the setup of MiniCACTUS installed in the beam area.

The \qtyproduct{10x10}{\centi\meter} PCB holding the sensor and ancillary electronics (regulators, buffers, etc) is attached to an aluminum structure. 
Behind the sensor, there is a scintillator and Photo Multiplier Tube (PMT - Hamamatsu 11934) assembly, used also in-lab tests. The PMT is biased at \SI{900}{V}. Its time resolution has been measured in the lab using cosmic events and found to be \SI{50}{ps} for MIPs.

The sensor is configured through a GPAC board (General Purpose Analog Card developed by the University of Bonn), attached to a Raspberry Pi, which acts as a general control gate to the setup through its Ethernet interface. The GPAC cards feature SPI and I2C interfaces to various DACs that are used to configure the MiniCACTUS. The high voltage is applied to the sensor by a Keithley 2410 sourcemeter. Data acquisition is performed using a high end digital oscilloscope (LeCroy WaveRunner 8254 or similar), controlled remotely via VNC (Virtual Network Computing). The sampling rate used by the oscilloscope was 10~GSPS for 4 channels. The high voltage, whose value was typically between \SI{-100}{V} and \SI{-500}{V} was controlled and monitored remotely via the Raspberry Pi attached to Ethernet. The whole setup, including the MiniCACTUS sensors, was running at room temperature and there was no temperature control.

The aluminum structure is itself attached to a movable table. After installation, a first coarse alignment of the sensor with respect to the beamline is done mechanically. A more precise alignment is done afterward, using the RD-51 tracker put in coincidence with the MiniCACTUS analog monitoring signal (AmpOut). Since the RD-51 tracker acceptance is of the order of \qtyproduct{10x10}{\centi\meter}, which is much larger than the MiniCACTUS, it is straightforward to locate precisely the position of the sensor with respect to the RD-51 beam telescope and to optimize the MiniCACTUS position with respect to the beam axis, whose direction is known precisely with respect to the beamline telescope axis.

Data presented in this paper \textcolor{\txtc}{were} taken using muon beams only. The MiniCACTUS acquisition by the LeCroy oscilloscope was triggered in the presence of an analog signal with sufficient amplitude (a few tens of mV) on the AmpOut analog monitoring output. The signals from the MiniCACTUS, AmpOut and DigOut, from the PMT, and from one of the beamline MCPPMTs are all digitized by the oscilloscope. Since the setup provides three time measurements (the MiniCACTUS, the PMT, and the MCPPMT), it is possible to measure simultaneously the time resolution of all three devices.

\section{Analysis of results}
\label{Analysis_of_results}
\begin{figure}[htbp]
\centerline{\includegraphics[width=3.7in]{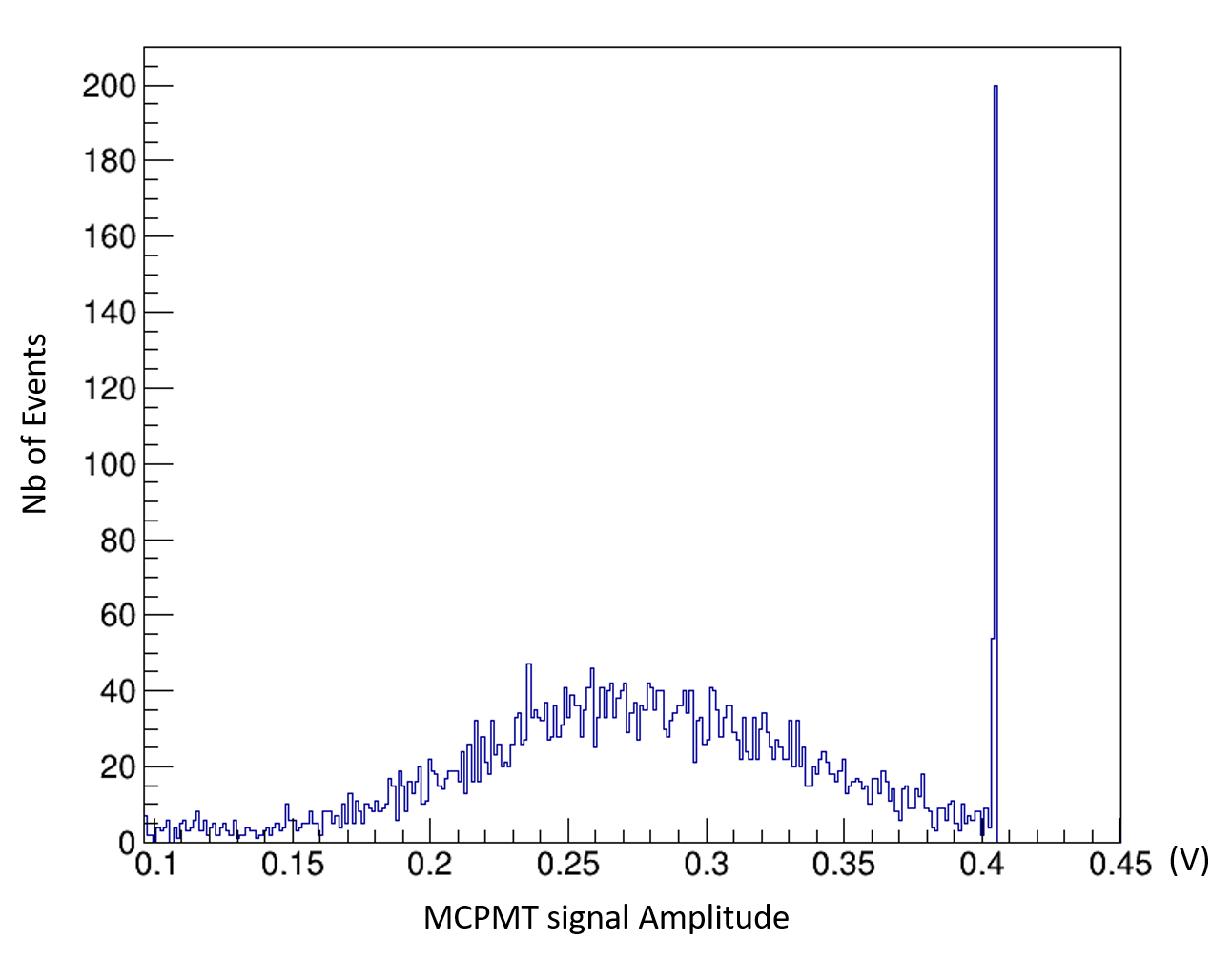}}
\caption{MCPPMT signals amplitude (V) distribution for muons beam.}
\label{Ana1}
\end{figure}

The result presented below corresponds to boards 5 and 6, which are equipped with two different \SI{185}{\micro m}-thick sensors. FE parameters were first studied and optimized in the lab using a $^{90}$Sr source. During the test beam, the optimized FE parameters were applied to the pixel 8 with a \SI{-500}{V} bias voltage. A hard threshold on the MCPPMT signals was set to select muon producing a signal amplitude above \SI{150}{mV}. The energy spectrum obtained from the MCPPMT is shown in Fig.~\ref{Ana1}. \textcolor{\txtc}{The rightmost bin on this figure, with highest number of events corresponds to events where the measured amplitude exceeds the maximum range of the digital scope setting for this channel}. The MCPPMT time resolution measured by the RD-51 team using such devices is below \SI{10}{ps} \cite{mcp}.

\begin{figure}[htbp]
\centerline{\includegraphics[width=3.6in]{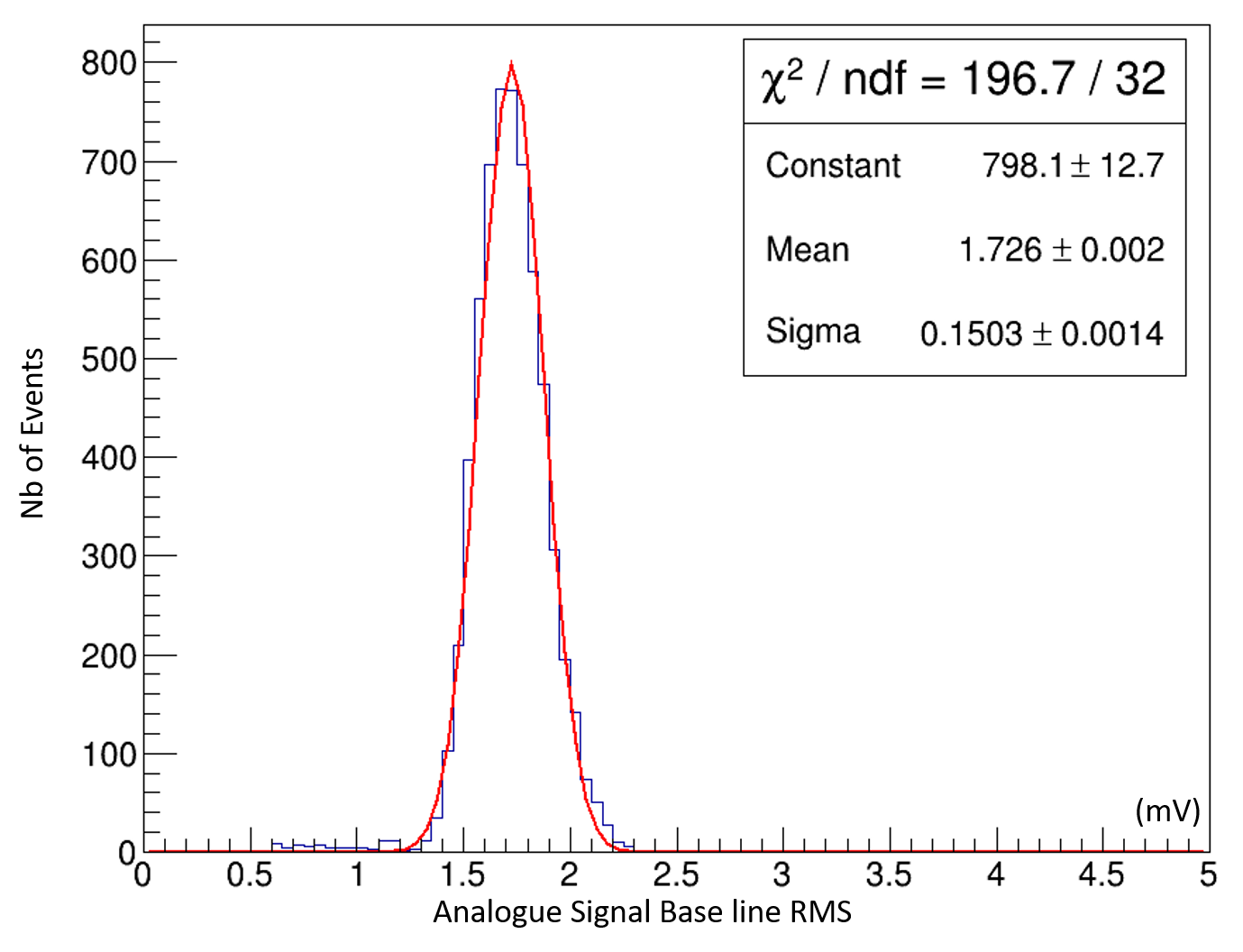}}
\caption{Analogue signal base line RMS distribution of MiniCACTUS (pixel 8 for a \SI{185}{\micro m}-thick sensor with nominal FE settings at \SI{-500}{V}) \textcolor{\txtc}{using a Gaussian fit}.}
\label{Ana2}
\end{figure}

We estimated a total noise of \SI{1.73}{mV} by measuring the analog signal base line RMS value \textcolor{\txtc}{using a gaussian fit} (see Fig.~\ref{Ana2}). This value is consistent with the measurements done previously in the laboratory and is estimated to correspond to \SI{170}{e^-} ENC.

\begin{figure}[htbp]
\centerline{\includegraphics[width=3.6in]{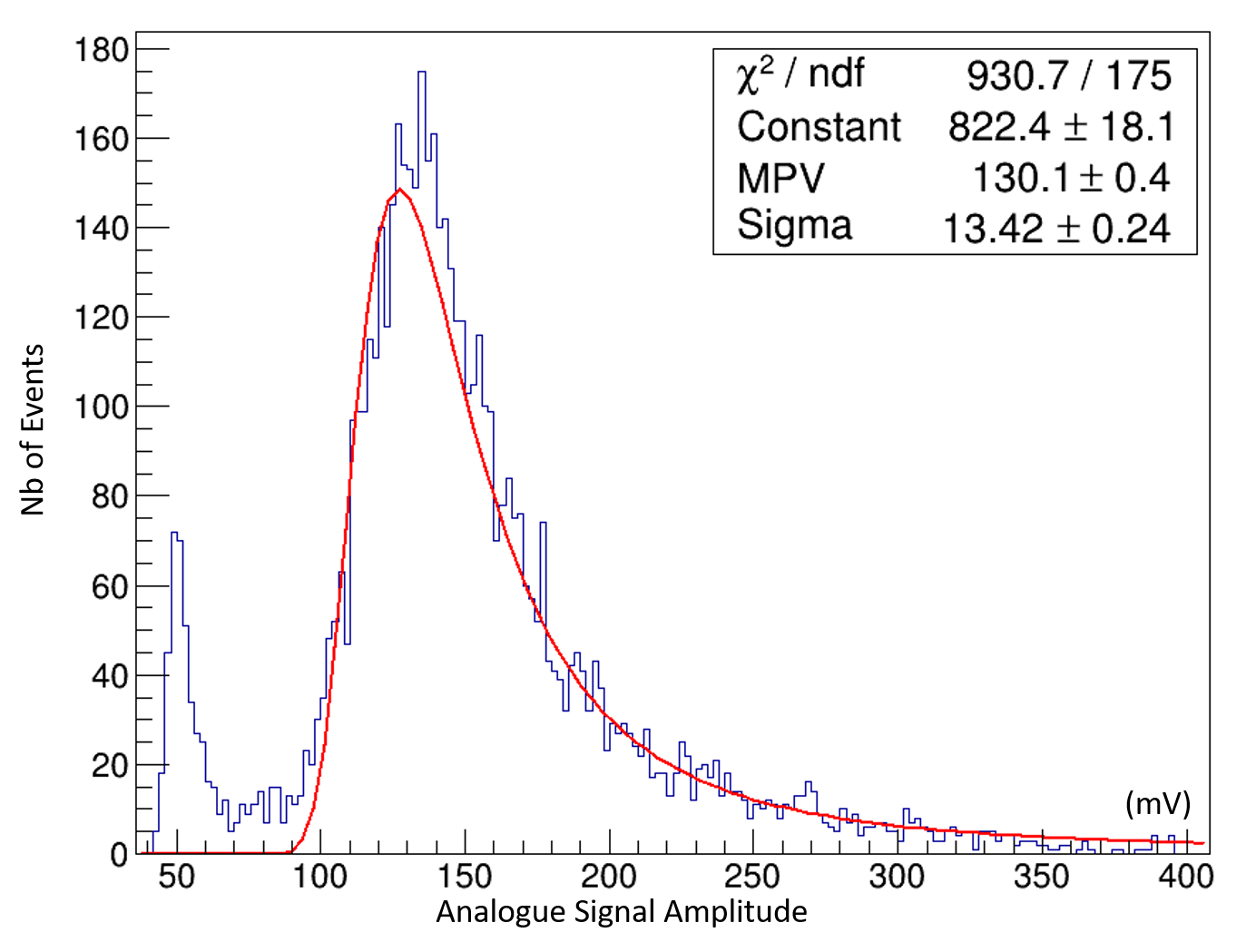}}
\caption{Analogue signal amplitude distribution of MiniCACTUS for muons beam (pixel 8 for a \SI{185}{\micro m}-thick sensor with nominal FE settings at \SI{-500}{V}) \textcolor{\txtc}{using a Landau distribution fit. Parameters of the Landau curve are given on the picture.}}
\label{Ana3}
\end{figure}

\begin{figure}[htbp]
\centerline{\includegraphics[width=3.8in]{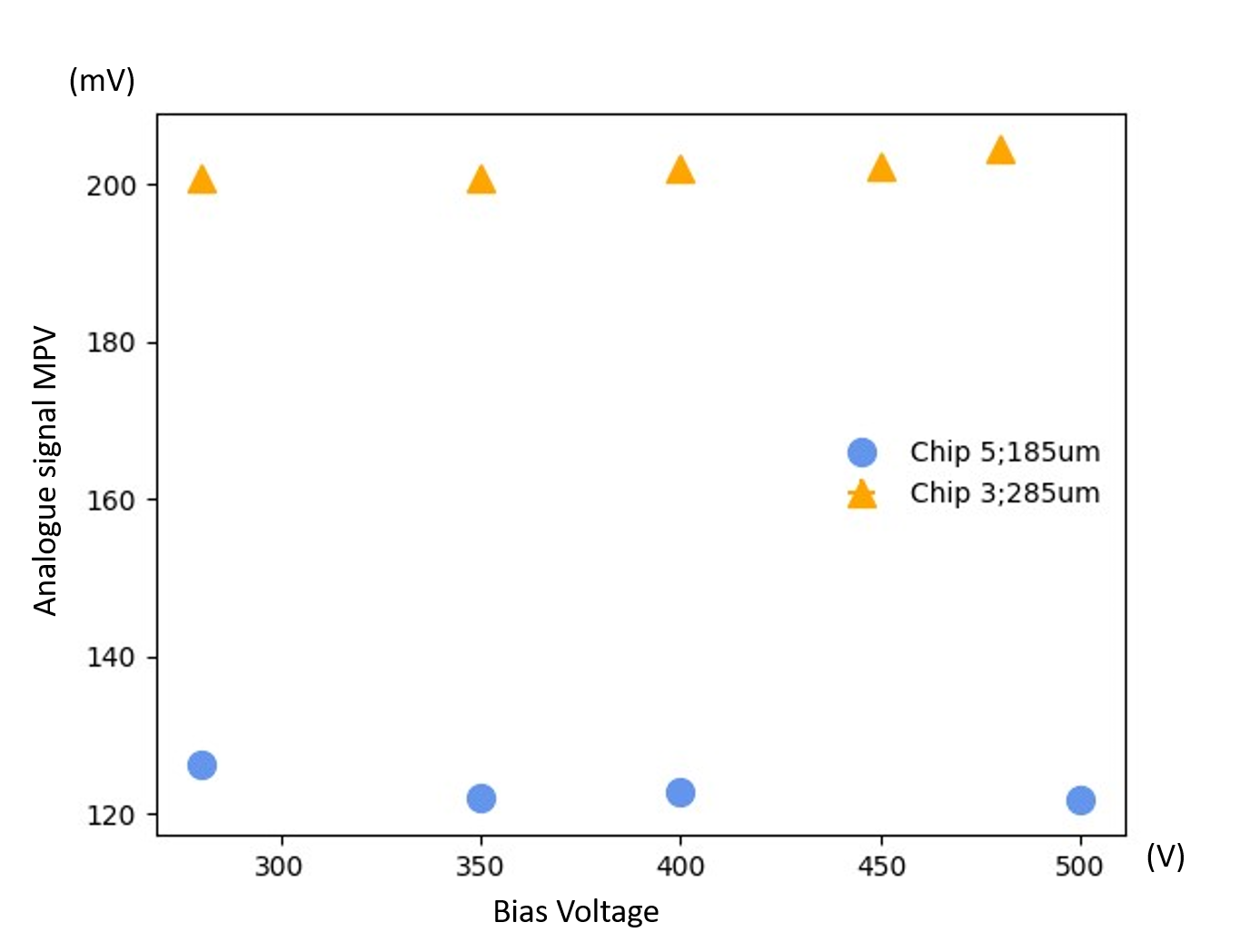}}
\caption{MiniCACTUS analog signal MPV as a function of HV bias voltage for chip 5 (\SI{185}{\micro m}) and chip 3 (\SI{285}{\micro m}).}
\label{Ana9}
\end{figure}

\begin{figure}[htbp]
\centerline{\includegraphics[width=3.7in]{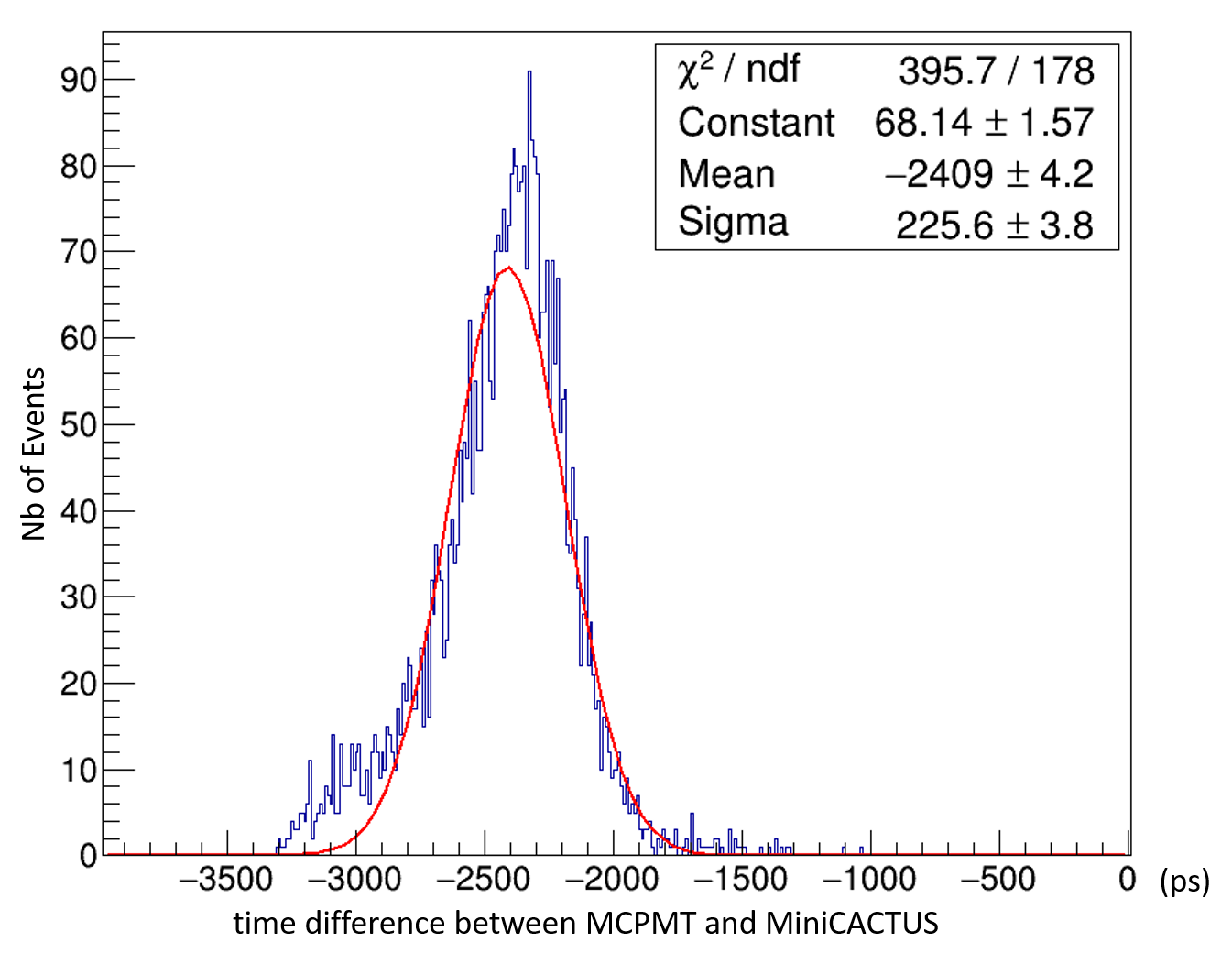}}
\caption{Time difference between MCPPMT and MiniCACTUS digital signals before TW correction for muons beam (pixel 8 for a \SI{185}{\micro m}-thick sensor with nominal settings at \SI{-500}{V}) \textcolor{\txtc}{fitted to a gaussian distribution. Parameters of the gaussian are given on the picture.}}
\label{Ana4}
\end{figure}

\begin{figure}[htbp]
\centerline{\includegraphics[width=3.6in]{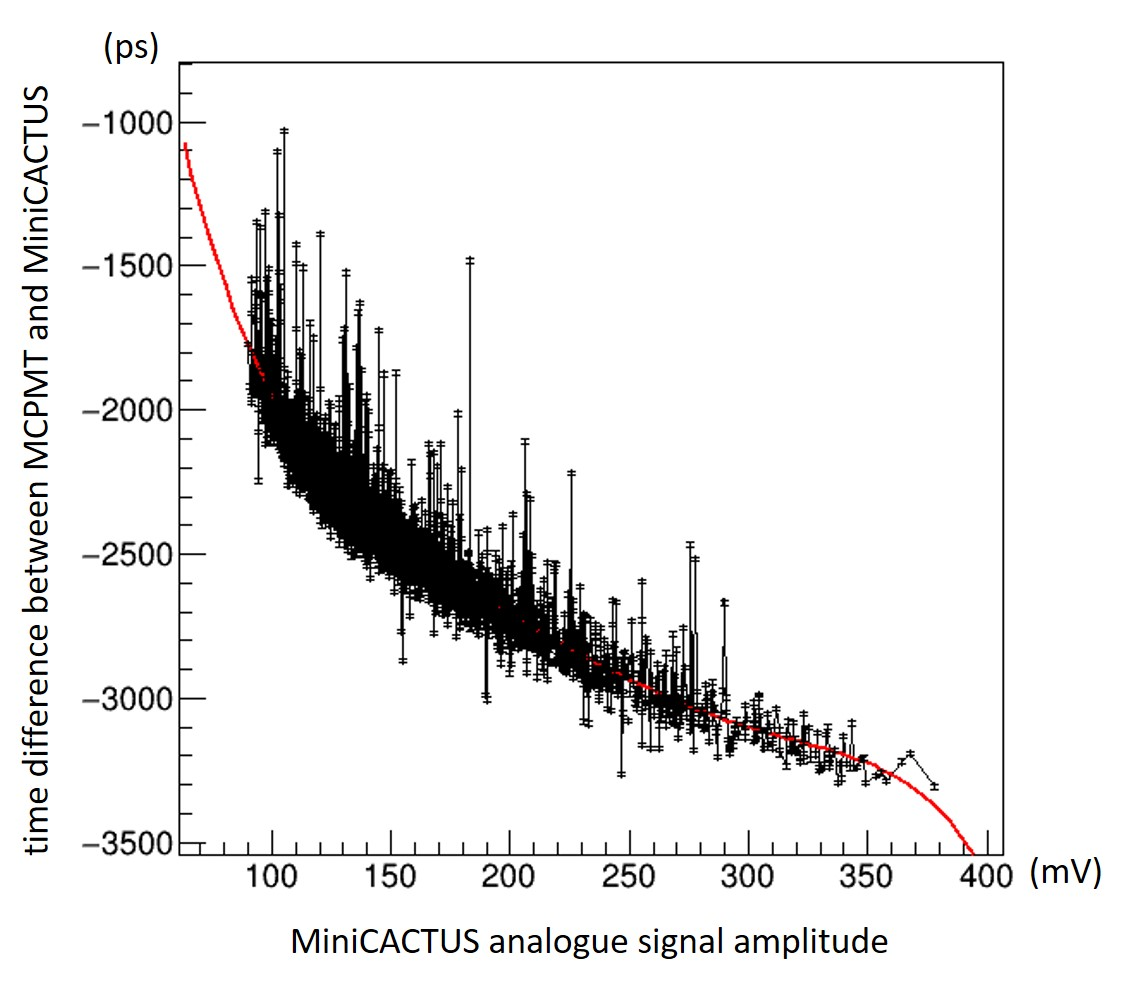}}
\caption{Time difference between MCPPMT and MiniCACTUS digital signals versus MiniCACTUS analogue signal amplitude.}
\label{Ana5}
\end{figure}

\begin{figure}[htbp]
\centerline{\includegraphics[width=3.6in]{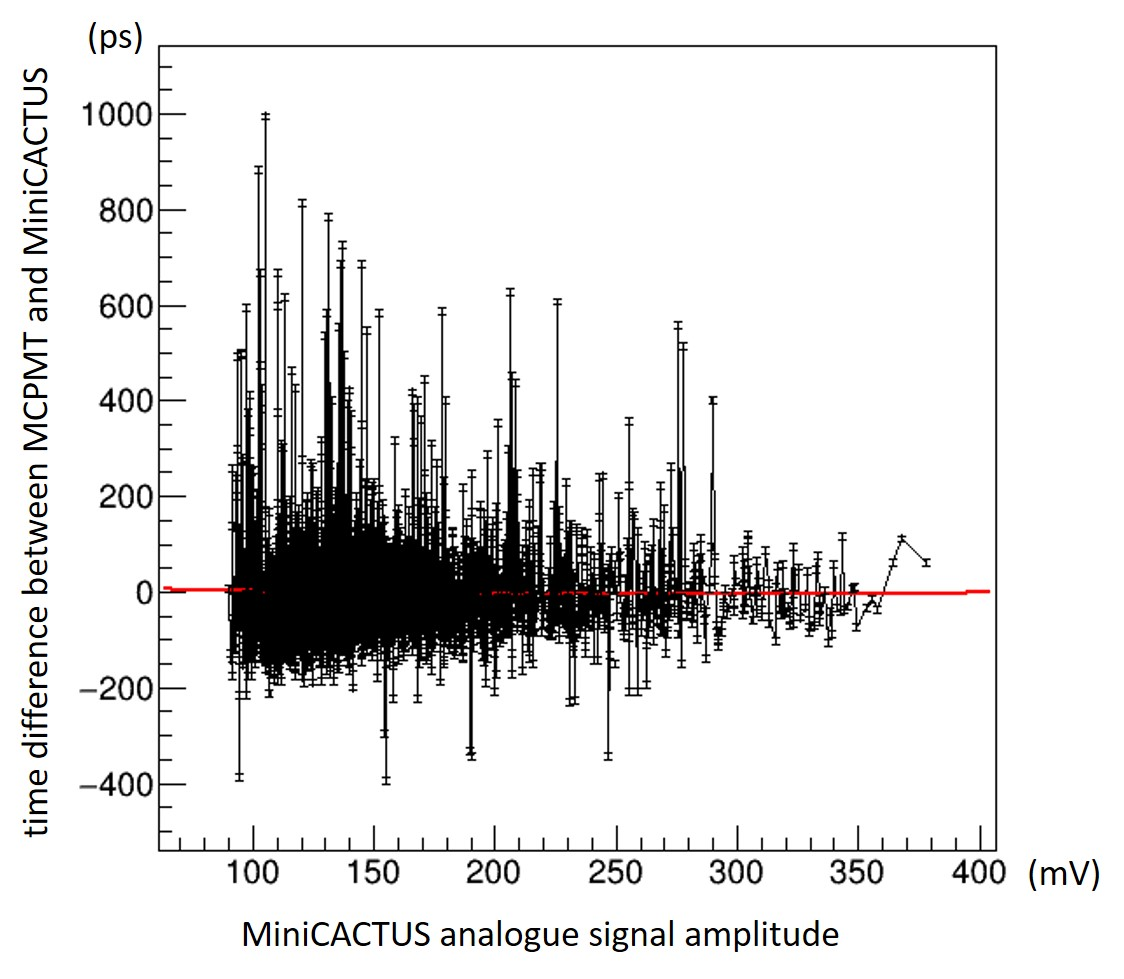}}
\caption{Time difference distribution between MCPPMT and MiniCACTUS digital signals versus MiniCACTUS analogue signal amplitude after TW correction.}
\label{Ana6}
\end{figure}

\begin{figure}[htbp]
\centerline{\includegraphics[width=3.8in]{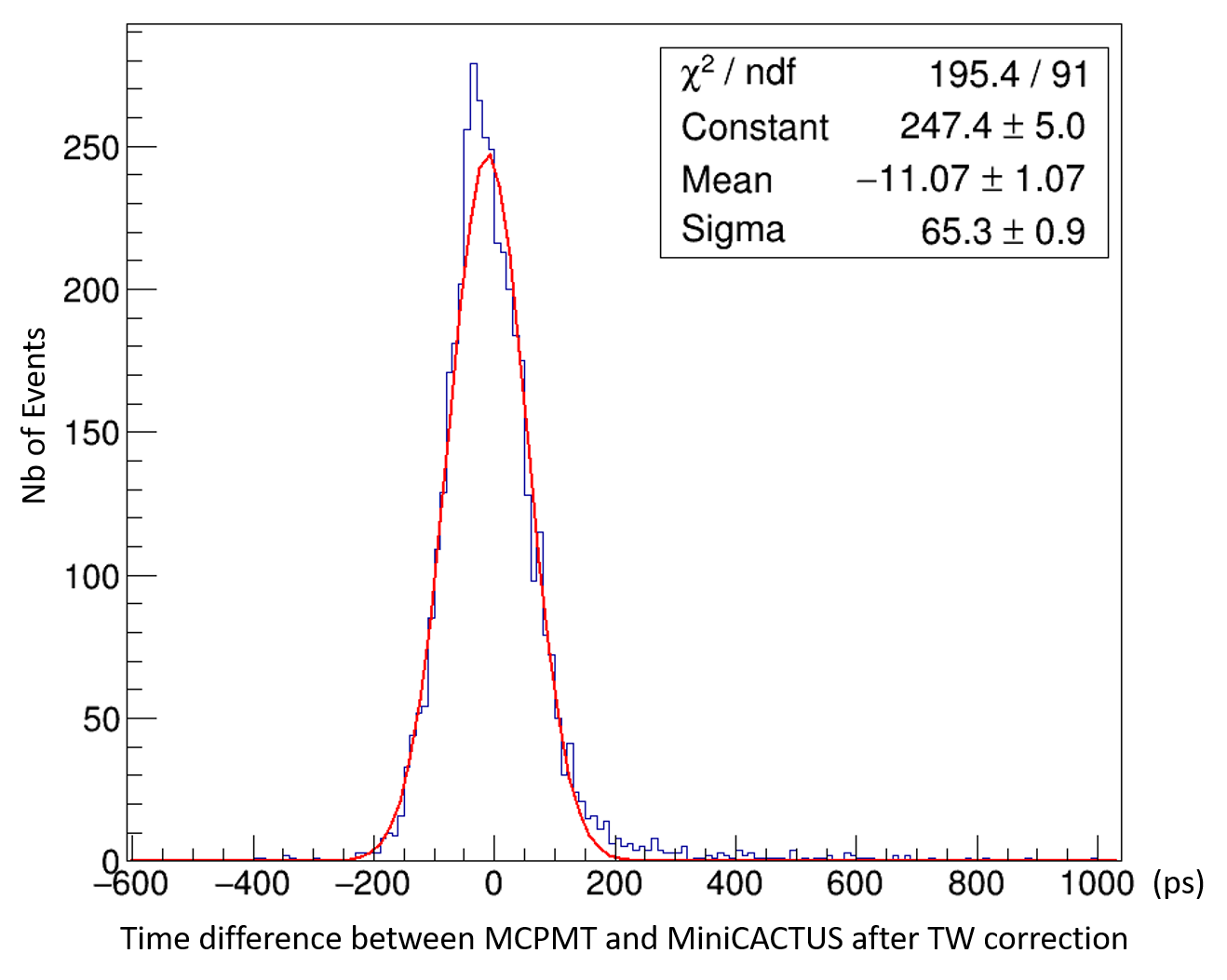}}
\caption{Time difference distribution between MCPPMT and MiniCACTUS digital signals after TW correction (pixel 8 for a \SI{185}{\micro m}-thick sensor with nominal FE settings at \SI{-500}{V}) \textcolor{\txtc}{fitted to a gaussian distribution. Parameters of the gaussian are given on the picture.}}
\label{Ana7}
\end{figure}

\textcolor{\txtc}{The Most Probable Value (MPV) of the analog signal obtained by a Landau distribution fit,} in the presence of particles on the detector is \SI{130}{mV} (Fig.~\ref{Ana3}). Therefore, the SNR is of the order of 75. This result does not depend on the high voltage applied to the chip  as shown in Fig.~\ref{Ana9} and is consistent with the fact that the sensor is fully depleted in the high voltage range we have explored in test-beam (\qtyrange{-250}{-500}{V}). To further select the good events, we request the presence of a MiniCACTUS digital signal DigOut, indicating that the internal discriminator has fired. The threshold was set at 15 mV above the baseline voltage corresponding to about one tenth of the MPV amplitude at the CSA output. \textcolor{\txtc}{Both the measurement of the time of arrival of MCPPMT signal and the measurement of the time of arrival of the MiniCACTUS are affected by time jitter. For each of these signals, the time jitter depends upon the risetime and the SNR. Hence, the r.m.s. of the distribution of the time difference between the time of arrival of the MCPPMT and the time of arrival of the MiniCACTUS digital signal is the quadratic sum of the time resolution of the MPCCPMT and of the MiniCACTUS. In practice, When extracting the time resolution from the distribution of the time difference between MCPPMT and the MiniCACTUS digital signal, the below 10 ps contribution from the MCPPMT is negligible}. The MiniCACTUS time resolution measured before Time Walk (TW) correction is \SI{225.6}{ps} (Fig.~\ref{Ana4}). Since we use a constant threshold leading edge discriminator, \textcolor{\txtc}{the difference between the time of arrival of the particle and the time at which the MiniCACTUS discriminator fires depends upon the amplitude of the analog signal. This is the time walk effect \cite{Cartiglia}. As a consequence, }the time difference between the MCPPMT signal and the MiniCACTUS depends on the analog signal amplitude. The amplitude (measured using the analog monitoring signal AmpOut) dependence upon the difference between MCPPMT time of arrival and MiniCACTUS time of arrival is corrected offline using a fifth-order polynomial (Fig.~\ref{Ana5}). After TW correction (Fig.~\ref{Ana6}), the MiniCACTUS time resolution is \SI{65.3}{ps} (Fig.~\ref{Ana7}). 


\begin{figure}[htbp]
\centerline{\includegraphics[width=3.8in]{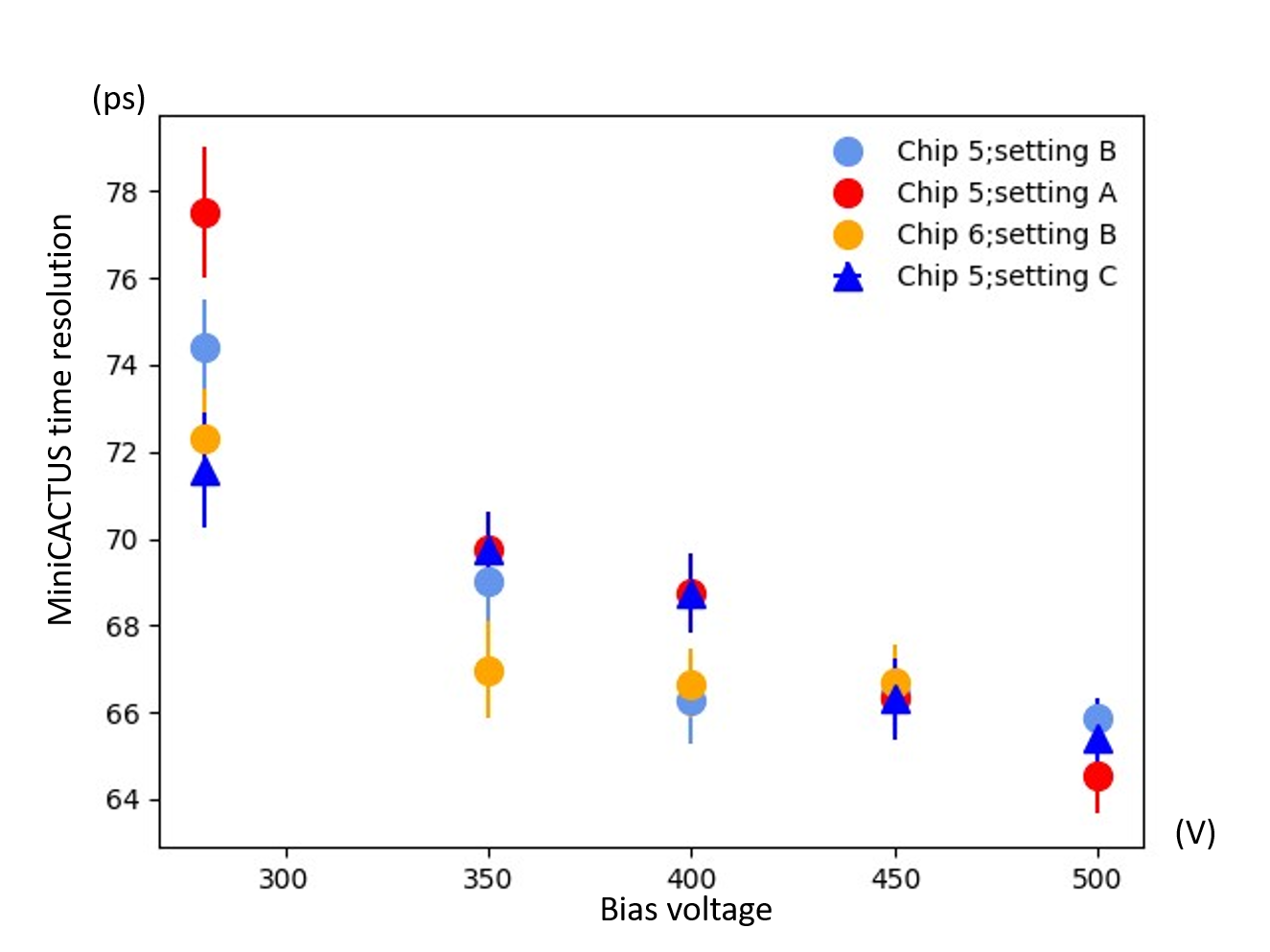}}
\caption{MiniCACTUS time resolution as a function of HV bias voltage for board 5 and 6 (Setting A: default FE settings; Setting B: optimized bandwidth and peaking time; Setting C: reduced ToT).}
\label{Ana8}
\end{figure}

As explained in Section II, the FE has been designed originally with a peaking time of \SI{\sim 1}{ns} and a Time-over-Threshold (ToT) of \SI{\ge 30}{ns} for a detector capacitance of \SI{1.5}{pF}. For \SI{185}{\micro m}-thick sensors, an analysis of the observed pulse shapes from MIPs indicates that the charge collection time is \SI{\ge 4}{ns}. So with default SC settings (called \textit{setting A} here), the peaking time and bandwidth of the FE are not optimized for these conditions. Unfortunately, there is no possibility to tune the peaking time of the FE for this chip directly, but by modifying the FE biasing parameters (see Fig. ~\ref{FE}), it is possible to adjust slightly the bandwidth and peaking time of the CSA with negligible increase of the power consumption. The modified FE control parameters are:
\textit{BLRES} (setting of the equivalent resistance of the C-R baseline restorer),
\textit{IFB} (setting of the feedback current of the CSA),
and \textit{IFOLL} (bias current of the output source follower stage of the CSA, not detailed in Fig. ~\ref{FE}).
The \textcolor{\txtc}{light blue} points of Fig.~\ref{Ana8} have been obtained by increasing \textit{IFOLL} and \textit{BLRES} (\textit{Setting B)}. \textcolor{\txtc}{A second sensor has also been tested in these conditions giving even better timing results (yellow points).} It is also possible to reduce the ToT by increasing \textit{IFB} (\textit{Setting C}) to have the ToT close to~\SI{25}{ns}. This feature is particularly important to minimize pile up effects in high luminosity collider environments like LHC. The plateau observed at high bias values could be \textcolor{\txtc}{induced by the intrinsic fluctuations of the magnitude and position of the charge deposits (Landau fluctuations, \cite{PDG})} for this sensor thickness. \textcolor{\txtc}{The bias conditions of Fig. ~\ref{Ana8} have been summarized in Table~\ref{dac-setting}.}

\begin{table}[h]
\color{\txtc}
    \centering
    \caption{\textcolor{\txtc}{List of Global Current Setting DACs of MiniCACTUS}}
    \label{dac-setting}
    \setlength{\tabcolsep}{.2cm}
    \begin{tabular}{lm{3.1cm}@{\hspace{.2cm}}S@{\hspace{.1cm}}S@{\hspace{.1cm}}S@{\hspace{.1cm}}m{0.5cm}}
               &                                                               & \multicolumn{3}{c}{\textbf{Settings}} &                                                            \\ \midrule
\textbf{DAC}   & \textbf{Definition}                                           & \textbf{A}              & \textbf{B}              & \textbf{C}             & \textbf{unit} \\ \midrule
BLRES          & Baseline restorer transistor control current                  & 1.0                     & 21.0                    & 1.0                    & nA   \\ \noalign{\smallskip}
IFOLL          & CSA Source Follower stage bias current                        & 2.5                     & 7.5                     & 10.0                   & µA   \\ \noalign{\smallskip}
FB             & CSA feedback current                                          & 12.0                    & 12.0                    & 30.0                   & nA   \\ \noalign{\smallskip}
IPBIAS            & CSA gain stage total current                                  & 800.0                   & 800.0                   & 800.0                  & µA  \\ \bottomrule
\end{tabular}
\end{table}

\begin{table}[htbp]
\color{\txtc}
    \centering
    \caption{\textcolor{\txtc}{Measured SNRs for Chip 5 (185~µm) at 500~V}}
    \label{Chip5-SNR}
    \begin{tabular}{@{}cccc@{}}
\toprule
\textbf{Pixel} & \textbf{Pitch}    & \textbf{SNR} & \textbf{Err SNR} \\
               & mm $\times$ mm     &              &                   \\ \midrule 
8              & 1.0 $\times$ 0.5   & 75.5         & 0.2              \\
7              & 1.0 $\times$ 1.0   & 58.9         & 0.2              \\
5              & 0.05 $\times$ 0.15   & 77.7         & 0.5              \\ \bottomrule
    \end{tabular}
\end{table}
 
\begin{table}[htbp]
\color{\txtc}
    \centering
    \caption{\textcolor{\txtc}{Measured SNRs for Chip 3 (285~µm) at 450~V}}
    \label{Chip3-SNR}
    \begin{tabular}{@{}ccSc@{}}
\toprule
\textbf{Pixel} & \textbf{Pitch}     & \textbf{SNR}  & \textbf{Err SNR} \\
               & mm $\times$ mm     &               &                   \\ \midrule 
8              & 1.0 $\times$ 0.5   & 103.1         & 0.2              \\
7              & 1.0 $\times$ 1.0   & 95.2          & 0.3              \\
5              & 0.05 $\times$ 0.15   & 97.5          & 1.6              \\ \bottomrule
    \end{tabular}
\end{table}

During these test beam campaigns, \SI{285}{\micro m} and \SI{85}{\micro m}-thick sensors have also been tested. Despite more charges collected with muons, the time resolution of \SI{285}{\micro m} sensors was measured to be \SI{75}{ps} for the same pixel 8. The distribution of the signal height, which is proportional to the total collected charge, follows a Landau distribution, with its MPV corresponding approximately to a collected charge $q_{tot}=\SI{80}{e^-} \times d$, where $d$ is the depleted thickness, in microns. Fig.~\ref{Ana9} shows the measured MPVs for two different sensors, \SI{185}{\micro m} and \SI{285}{\micro m}, as a function of the bias voltage. The corresponding SNRs are \SI{\sim 75}{} and \SI{\sim 100}{} respectively. The worsening of the time resolution with the increase of thickness could be explained by the expected increase of the Landau fluctuations with increased thickness. Concerning the \SI{85}{\micro m} sensors, it was not possible to get a clean Landau distribution with muons due to the limited SNR and also strong couplings from the digital buffers (explained at the end of Section II) at low thresholds. 

\textcolor{\txtc}{The small test pixels show also good SNR (Tables~\ref{Chip5-SNR}~\&~\ref{Chip3-SNR})} but worse timing resolution due to charge sharing at these thicknesses. They need probably thinner substrates or epi-wafers to be exploited \textcolor{\txtc}{for timing}.

\section{Conclusions}
MiniCACTUS, a monolithic timing sensor prototype has been designed and fabricated in the LFoundry \SI{150}{nm} CMOS process. After in-lab tests, the backside-processed \textcolor{\txtc}{\SI{85}{\micro m}, \SI{185}{\micro m} and \SI{285}{\micro m}-thick sensors (active part thicknesses)} have been tested extensively in testbeam with MIPs at CERN SPS. The best results have been obtained with \SI{185}{\micro m} sensors : 
A time resolution of \SI{65.3}{ps} has been measured with 0.5\,mm$\times$ 1.0\,mm pixels biased at \SI{-500}{V} with muons. The measurement includes already the on-chip FE and discriminator. This time resolution has been measured consistently over several testbeam campaigns and with several sensors. The power consumption under these conditions is \SI{300}{\milli\watt\per\centi\meter\squared}, which is compatible with the requirements of large high energy physics collider experiments.
This prototype is an important step towards a fully monolithic large size sensor.

\section*{Acknowledgment}

The authors would like to acknowledge the RD-51 team for their kindful help during the testbeam campaigns at SPS-CERN.


\begin{thebibliography}{00}
\bibitem{HGTD} C. Allaire \textit{et al.}, ``Beam test measurements of Low Gain Avalanche Detector single pads and arrays for the ATLAS High Granularity Timing Detector,''~\textit{J. Intsrum.}, vol. 13, June 2018, Art. no. P06017, doi: https://doi.org/10.1088/1748-0221/13/06/p06017.

\bibitem{MTD} CMS Collaboration, ``A MIP Timing Detector for the CMS Phase-2 Upgrade,'' Technical Design Report, CERN, Geneva, 2019, [Online]. Avalaible: https://cds.cern.ch/record/2667167.

\bibitem{HGCAL} CMS Collaboration, ``The Phase-2 Upgrade of the CMS Endcap Calorimeter,'' Technical Design Report, CERN, Geneva, 2017,  [Online]. Avalaible: https://cds.cern.ch/record/2293646.

\bibitem{Apresyan} A. Apresyan \textit{et al.}, ''Studies of uniformity of \SI{50}{\micro m} low-gain avalanche detectors at Fermilab test beam,''~\textit{Nucl. Instrum. Methods Phys. Res. A, Accel. Spectrom. Detect. Assoc. Equip.}, vol. 895, pp. 158--172, July 2018, doi: https://doi.org/10.1016/j.nima.2018.03.074.

\bibitem{ibl-tdr} ATLAS Collaboration, ''ATLAS Insertable B-Layer Technical Design Report'', CERN-LHCC-2010-013, ATLAS-TDR-19, Sep. 2010. [Online]. Available: https://cds.cern.ch/record/1291633

\bibitem{itk-tdr} ATLAS Collaboration, ''Technical Design Report for the ATLAS Inner Tracker Pixel Detector'', CERN-LHCC-2017-021, ATLAS-TDR-030, Sep. 2017. [Online]. Available: https://cds.cern.ch/record/2285585

\bibitem{Peric} I. Peric, ``A novel monolithic pixelated particle detector implemented in high-voltage CMOS technology,''~\textit{Nucl. Instrum. Methods Phys. Res. A, Accel. Spectrom. Detect. Assoc. Equip.}, vol. 582, no. 3, pp. 876--885, December 2007, doi: https://doi.org/10.1016/j.nima.2007.07.115.

\bibitem{Chen} Z. Chen \textit{et al.}, ``Test results of irradiated CMOS pixel circuits in 150 nm CMOS technology for the ATLAS Inner Tracker Upgrade,''~\textit{PoS}, vol. 343 (TWEPP2018), July 2019, doi: https://doi.org/10.22323/1.343.0156.

\bibitem{Barbero} M. Barbero \textit{et al.}, ``Radiation hard DMAPS pixel sensors in 150nm CMOS technology for operation at LHC,''~\textit{J. Instrum.}, vol. 15, May 2020, Art. no. P05013, doi: https://doi.org/10.1088/1748-0221/15/05/P05013.

\bibitem{cactus} Y. Degerli, F. Guilloux, C. Guyot, J.P. Meyer, A. Ouraou, P. Schwemling, A. Apresyan, R. Heller, M. Mohd, C. Pena, S. Xie and T. Hemperek, ''CACTUS: a depleted monolithic active timing sensor using a CMOS radiation hard technology,''~\textit{J. Instrum.}, vol. 15, no. 6, Art. no. P06011, June 2020, doi: https://doi.org/10.1088/1748-0221/15/06/P06011.

\bibitem{minicactus} Y. Degerli, F. Guilloux, T. emperek, J.P. Meyer, P. Schwemling, ''MiniCACTUS: Sub-100 ps timing with depleted MAPS,''~\textit{Nucl. Instrum. Methods Phys. Res. A, Accel. Spectrom. Detect. Assoc. Equip.}, vol. 1039, no. 167022, Sep. 2022, doi: https://doi.org/10.1016/j.nima.2022.167022.

\bibitem{lfmonopix2} I. Caicedo \textit{et al.}, ''Development and testing of a radiation-hard large-electrode DMAPS design in a 150 nm CMOS process,''~\textit{Nucl. Instrum. Methods Phys. Res. A, Accel. Spectrom. Detect. Assoc. Equip.}, vol. 1040, no. 167224, Oct. 2022, doi: https://doi.org/10.1016/j.nima.2022.167224.

\bibitem{Mandic} I. Mandic \textit{et al.}, ''Charge-collection properties of irradiated depleted CMOS pixel test structures''~\textit{Nucl. Instrum. Methods Phys. Res. A, Accel. Spectrom. Detect. Assoc. Equip.}, vol. 903, Sep. 2018, doi: https://doi.org/10.1016/j.nima.2018.06.062.

\bibitem{mcp} J. Bortfeldt \textit{et al.}, ''Timing performance of a micro-channel-plate photomultiplier tube,''~\textit{Nucl. Instrum. Methods Phys. Res. A, Accel. Spectrom. Detect. Assoc. Equip.}, vol. 960, no. 163592, April 2020, doi: https://doi.org/10.1016/j.nima.2020.163592.

\bibitem{Cartiglia} N. Cartiglia \textit{et al.}, ''Performance of ultra-fst silicon detectors,''~\textit{J. Instrum.}, vol. 9, Art. no. C02001, February 2014, doi: https://doi.org/10.1088/1748-0221/9/02/C02001.

\bibitem{PDG} R.L. Workman \textit{et al.} ''The Review of Particle Physics 2022'', \textit{Prog. Theor. Exp. Phys.} 2022, 083C01 (2022), section : ''Passage of particles through matter''

\end{thebibliography}
\end{document}